\pdfoutput=1

\documentclass[aps,10pt,twocolumn,showpacs,superscriptaddress,footnoteinbib]{revtex4}
\usepackage{amsmath}
\usepackage{latexsym}
\usepackage{amssymb}
\usepackage{graphics,epstopdf}
\usepackage[colorlinks=true, citecolor=blue, urlcolor=blue ]{hyperref}

\begin{document}
\title{Local simulation of singlet statistics for restricted set of measurement}

\author{Ashutosh Rai}
\email{arai@bose.res.in}
\affiliation{S.N. Bose National Center for Basic Sciences, Block JD, Sector III, Salt Lake, Kolkata-700098, India}

\author{MD Rajjak Gazi}
\email{rajjakgazimath@gmail.com}
\affiliation{Physics and Applied Mathematics Unit, Indian Statistical Institute, 203 B.T. Road, Kolkata-700108, India}

\author{Manik Banik}
\email{manik11ju@gmail.com}
\affiliation{Physics and Applied Mathematics Unit, Indian Statistical Institute, 203 B.T. Road, Kolkata-700108, India}

\author{Subhadipa Das}
\email{sbhdpa.das@bose.res.in}
\affiliation{S.N. Bose National Center for Basic Sciences, Block JD, Sector III, Salt Lake, Kolkata-700098, India}

\author{Samir Kunkri}
\email{skunkri@yahoo.com}
\affiliation{Mahadevananda Mahavidyalaya, Monirampore, Barrakpore, North 24 Parganas-700120, India}

\begin{abstract}
The essence of Bell's theorem is that, in general, quantum statistics cannot
be reproduced by a local hidden variable (LHV) model. This impossibility
is strongly manifested when statistics collected by measuring certain local
observables on a singlet state, violates the Bell inequality. In this work, we
search for local POVMs with binary outcomes for which an LHV model can
be constructed for a singlet state. We provide various subsets of observables
for which an LHV model can be provided for singlet statistics.
\end{abstract}
\maketitle

\section{Introduction}
A violation of the Bell-CHSH inequality \cite{bell} by statistics generated from local measurements performed on an entangled state shared between two spatially separated parties certifies such quantum state as nonlocal. The singlet state of two qubits (an EPR state) exhibits maximum nonlocality \cite{tsb} for proper choices of local observables. Although for pure entangled states the degree of nonlocality is in direct proportion to the entanglement content of a quantum state, this is, in general, not true for mixtures of entangled states \cite{gpp, ps}. Werner first gave the counterintuitive example of mixed entangled states (popularly known as Werner states) \cite{we}  whose statistics when subjected to projective measurements, can be generated by a local hidden variable (LHV) model. A similar example for tripartite entangled state which can be simulated by a local hidden variable model was first provided in a work by Toth and Acin \cite{acin}. A good review of research on hidden variable theories can be found in \cite{geno}.

Interestingly, Toner and Bacon \cite{tb} in the year 2003, gave a twist to earlier studies, by providing a model for singlet simulation which requires only 1 cbit of communication supplemented with local variables. Soon after, Cerf \emph{et al} \cite{cerf} showed that 1 \emph{nl}-bit (single PR-Box) is also sufficient for singlet simulation. Motivated by these works, recently, another model has been provided for singlet simulation which uses (possibly) signalling resource, namely $S^{p}$ correlations, which suggests a trade off relation between required communication and local randomness in measurement results \cite{gkar1,hall1}. Deggore \emph{et al.} \cite{deggo} could map the problem of simulating entangled states to distributed sampling problems.
A more through review of simulation of entangled state statistics from communication complexity point of view can be found in \cite{review}. Few other recent works \cite{hall2,hall3,gisin, manik} show that \emph{lack of free will} can also be considered as a resource for singlet simulation. There are also some efforts in solving the difficult problem of simulating  multipartite entanglement and non-maximally bipartite entangled states either by use of communication or by nonlocal (no-signaling) resources \cite{bru,ban}.  All these various approaches have been deepening our understanding about quantum correlation and its use as a physical resource in various information processing tasks.

As of providing local variable models for class of entangled states, in a seminal work in the year 2002, Barrett \cite{bar} generalized the work of  Werner \cite{we}, by construting a LHV model for any positive-operator-valued measurements at the expense of the weight associated with singlet in Werner state. Motivated by these works we, in this paper, pose the problem from opposite direction i.e. rather than weakening the (singlet) state we search for the class of (weakened) dichotomic observable (POVM) for which local model can be provided. In particular, here we provide the subset of the most general two outcome measurements represented by positive operator value measure (POVM) and present local models for singlet statistics generated from them. We provide some sets of local observable which are optimal for the protocol we have suggested. First we show that, if observable on any one side is sufficiently restricted (deviates from ideal projective measurement), resulting statistics for the singlet state has a local hidden variable model. Next, we provide another model which is symmetric in a sense that observable on both the sides are put to a similar type of restriction. Finally, we identify a more general set of observable for which  LHV models exists with some further restrictions.  Before we derive our results, in the followings section, we give a mathematical description of a general two-outcome POVMs.

\section{General two-outcome POVM}
Generalized quantum observables are described by POVMs \cite{bu}. For finite, say $n$, outcome measurements on a $d$-dimensional state space a POVM is a collection of selfadjoint operators $\{E_i\}$ acting on a complex Hilbert space $\mathbb{C}^{d}$ satisfying the conditions: (i) $0 \leq E_i \leq I $ for all $i$, and (ii) $\sum_{i} E_i=I$, where $i \in \{1,2,...,n\}$.
A measurement of such an observable $\{E_i\}$ on a quantum state $\rho$ results in any one of the $n$ possible outcomes; the probability of an occurrence of $i$-th outcome (termed as clicking of $i$-th \emph{effect}) is $Tr[\rho E_i]$.
 A subclass of these type of general measurement has an interesting physical interpretation as unsharp spin properties, introduced by P. Busch \cite{busch,busch1}.
 
\begin{figure}[!ht]
\begin{center}
\resizebox{8.0cm}{6.0cm}{\includegraphics{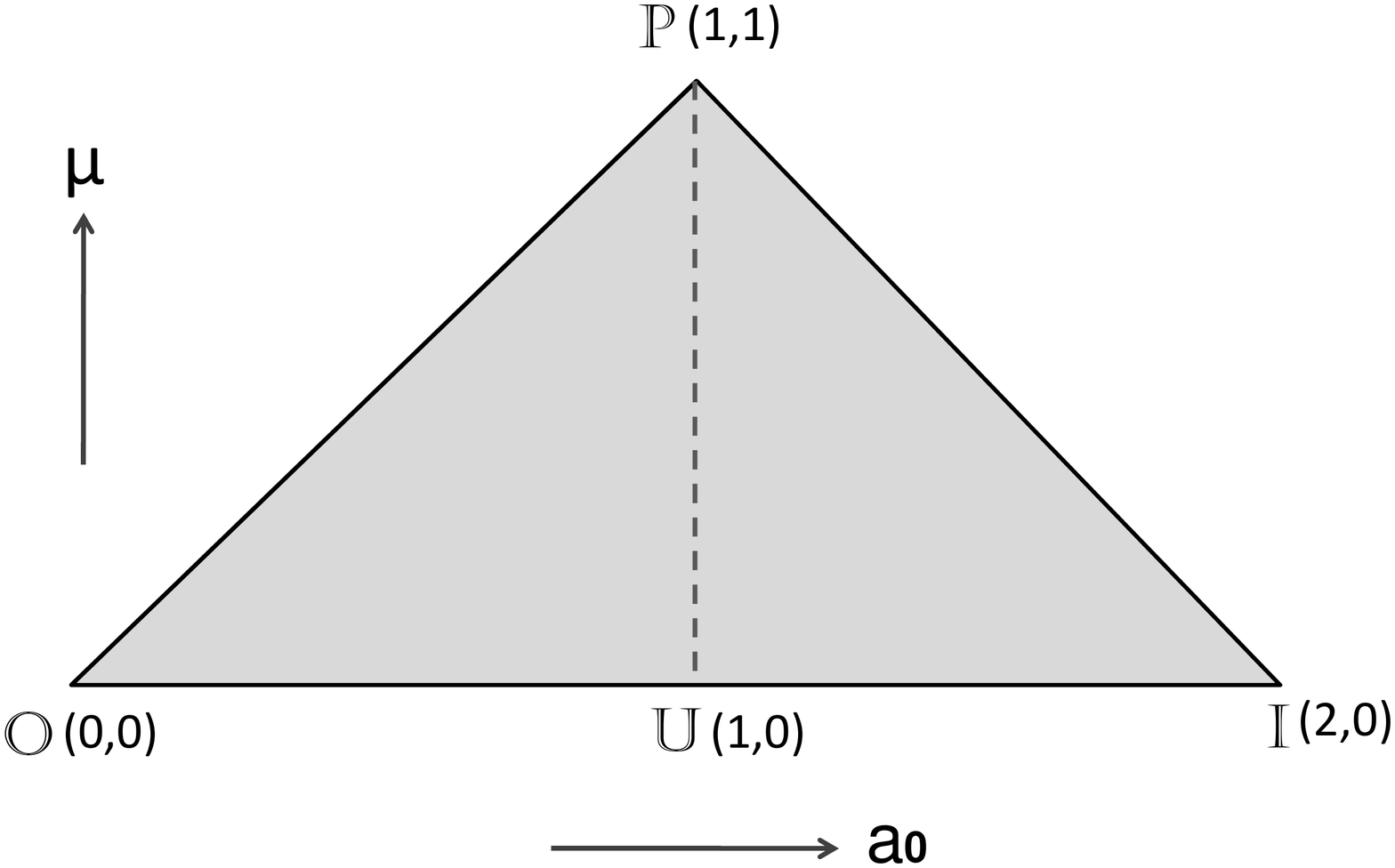}}
\end{center}
\caption{Parameter $a_0$ ($\mu$) varies along the horizontal (vertical) axis. Any point ($a_0$, $\mu$) laying in the shaded triangular region $\mathbb{POI}$ (together with an arbitrary parameter $\hat{a}$) determines a two-outcome POVM $\{E,I-E\}$. Points on the dashed line $\mathbb{PU}$ represent unsharp spin measurements. Point $\mathbb{P}(1,1)$ represent ideal projective measurements.\label{fig1}}
\end{figure}

In this work, we consider general two-outcome POVMs $\{E, I-E\}$ acting on $\mathbb{C}^{2}$ (state space of a qubit). Effect $E$ is characterized by some parameters, say, $a_0 \in \mathbb{R}$ (a scalar) and  $\vec{a}\in \mathbb{R}^{3}$ (a vector). We denote norm of $\vec{a}$ by $\mu$. Then, the selfadjoint property along with the condition $0 \leq E \leq I$ implies that $E$ can be expressed as\\
\begin{eqnarray}
E=\frac{1}{2}[a_0 I+ \mu\hat{a}\cdot\vec{\sigma}]\label{eq1} \\
0\leq a_0 \leq 2 \label{eq2}\\
0\leq \mu \leq\mbox{min}\{a_0,2-a_0\} \label{eq3}
\end{eqnarray}
where $\hat{a}\cdot\vec{\sigma}=a_x\sigma_x+a_y\sigma_y+a_z\sigma_z$. Then, the corresponding operator $I-E$ is also selfadjoint and satisfies the requirement $0 \leq I-E \leq I$. Thus, the Eq.(\ref{eq1}) along with the conditions (\ref{eq2}) and (\ref{eq3}), supplemented with an arbitrary direction $\hat{a}$, completely determine a two-outcome POVM $\{E,I-E\}$ acting on $\mathbb{C}^{2}$.
The region feasible for parameters $a_0$ and $\mu$ for defining such an effect $E\{a_0,\mu,\hat{a}\}$ is illustrated in Fig.(\ref{fig1}).

An interesting application of general two-outcome measurements considered here is in the study of spin properties of
spin-$\frac{1}{2}$ systems. In this context, P. Busch \cite{busch,busch1} first showed that a subclass of general two-outcome POVMs can be interpreted as measurement of unsharp-spin property of spin-$\frac{1}{2}$ particles. Under the condition of rotation covariance, parameters $\{a_0, \mu\}$ is decoupled from $\hat{a}$ which can then be interpreted as orientation of the measuring device. Further, condition of symmetry under a rotation $\pi$ of the measuring device gives $a_0=1$. Thus, effect operators for an unsharp spin observable is of the form $E^{\mu}_{\pm}(\hat{a})=\frac{1}{2}[I\pm\mu \hat{ a }\cdot\vec{\sigma}]$.
The spectral decomposition of positive operators $E^{\mu}_{\pm}(\hat{ a })$ is
\begin{equation}
~~~E^{\mu}_{\pm}(\hat{ a })=(\frac{1\pm\mu}{2})\frac{1}{2}[I+\hat{ a }.\vec{\sigma}]+
(\frac{1\mp\mu}{2})\frac{1}{2}[I-\hat{ a }.\vec{\sigma}] \nonumber
\end{equation}
where $\frac{1}{2}[I+\hat{ a }.\vec{\sigma}]$ and
$\frac{1}{2}[I-\hat{ a }.\vec{\sigma}]$ are one dimensional
spin projection operators on the Hilbert space $ \mathbb{C}^{2}$.
Now, the quantity $\frac{1+\mu}{2}$ ($\frac{1-\mu}{2}$) can be suitably interpreted as degree of reality
(unsharpness) of outcomes obtained from a spin measurement along direction $\hat{ a }$. From this representation it is clear that the POVM \{$E^\mu_+(\hat{ a }),E^\mu_-(\hat{ a })$\} is a smeared version of the projective measurement
\{$\frac{1}{2}[I+\hat{ a }.\vec{\sigma}],\frac{1}{2}[I-\hat{ a }.\vec{\sigma}]$\}---in case of projective measurements
the unsharp parameter $\mu=1$.

Another important property is that under suitable conditions two POVMs can be jointly measurable \cite{kraus}. Two POVMs of the form $\{E_1,I-E_1\}$ and $\{E_2,I-E_2\}$ are jointly measurable if there exits a four-outcome POVM $\{E_{12}, E_{\bar{1}2}, E_{1\bar{2}}, E_{\bar{1}\bar{2}} \}$ such that it can reproduce the correct marginals, i.e., $E_1 = E_{12} + E_{1\bar{2}}$ and $ E_2 = E_{12} + E_{\bar{1}2}$. For unsharp spin observable it has been shown that \cite{busch} (also see the review \cite{gkar}) two observables parameterized by, say, $(\mu_1, \hat{a}_1)$ and $(\mu_2, \hat{a}_2)$ are jointly measurable if and only if $\|\mu_1\hat{ a }_1+\mu_2\hat{ a }_2\|+\|\mu_1\hat{ a }_1-\mu_2\hat{ a }_2\|\leq 2$. On considering unsharp parameter for both the spin observables to be same i.e., $\mu_1 =\mu_2$, along with the fact $\|\hat{ a }_1+\hat{ a }_2\|+\|\hat{ a }_1-\hat{ a }_2\|\leq 2\sqrt{2}$ for any pair of unit vectors $\hat{ a }_1$ and $\hat{ a }_2$, it is easy to conclude that if the unsharp parameters $\mu_1 = \mu_2 \leq \frac{1}{\sqrt{2}}$ then joint measurement of unsharp spin property can be realized for any such pair of directions.

\section{LHV model for singlet statistics for two outcome POVMs}

Suppose, two spatially separated parties Alice and Bob share one qubit each from a singlet state $$\rho_{AB}=\frac{1}{4}[I\otimes I-\sigma_x\otimes \sigma_x-\sigma_y\otimes \sigma_y-\sigma_z\otimes \sigma_z].$$ Let Alice's (Bob's) observable be a most general two-outcome POVM $E_A[a_0,\mu_A,\hat{a}]$ ($E_B[b_0,\mu_B,\hat{b}]$), defined by Eq.(1). If the effect $E_{A(B)}$ clicks we denote the outcome by `\emph{yes}' otherwise `\emph{no}'. Then joint outcome probabilities are following:
\begin{eqnarray}
P^{AB}(yes,yes)&=& \frac{1}{4}[a_0 b_0-\mu_A\mu_B \hat{a}\cdot\hat{b}] \\
P^{AB}(yes,no)&=& \frac{1}{4}[a_0 (2-b_0)+\mu_A\mu_B \hat{a}\cdot\hat{b}] \nonumber\\
P^{AB}(no,yes)&=& \frac{1}{4}[(2-a_0) b_0+\mu_A\mu_B \hat{a}\cdot\hat{b}]\nonumber\\
P^{AB}(no,no)&=& \frac{1}{4}[(2-a_0)(2-b_0)-\mu_A\mu_B \hat{a}\cdot\hat{b}] \nonumber \label{eq4}
\end{eqnarray}

\subsection{Models for two-outcome measurements}
Violation of the Bell-CHSH inequality \cite{bell} implies that there can be no LHV model for the singlet statistics generated by projective measurements by both the parties. Therefore, the statistics of the singlet can have a LHV model only if general two outcome POVMs considered here are restricted (deviate from ideal projective measurements) in some way or the other.
Following Werner's local model for some mixed entangled states \cite{we}, we provide two LHV models for singlet state under certain restrictions on parameters of two outcome POVMs. In both type of models vectors $\hat{\lambda}=(\sin\theta \cos\phi,\sin\theta \sin\phi, \cos\theta)$ uniformly distributed over the unit sphere, are the local variables preshared between Alice and Bob.

\subsubsection{A fully biased model $\mathbb{M}_{fb}$:}
\begin{figure}[b]
\begin{center}$
\begin{array}{cc}
\resizebox{4.0cm}{2.5cm}{\includegraphics{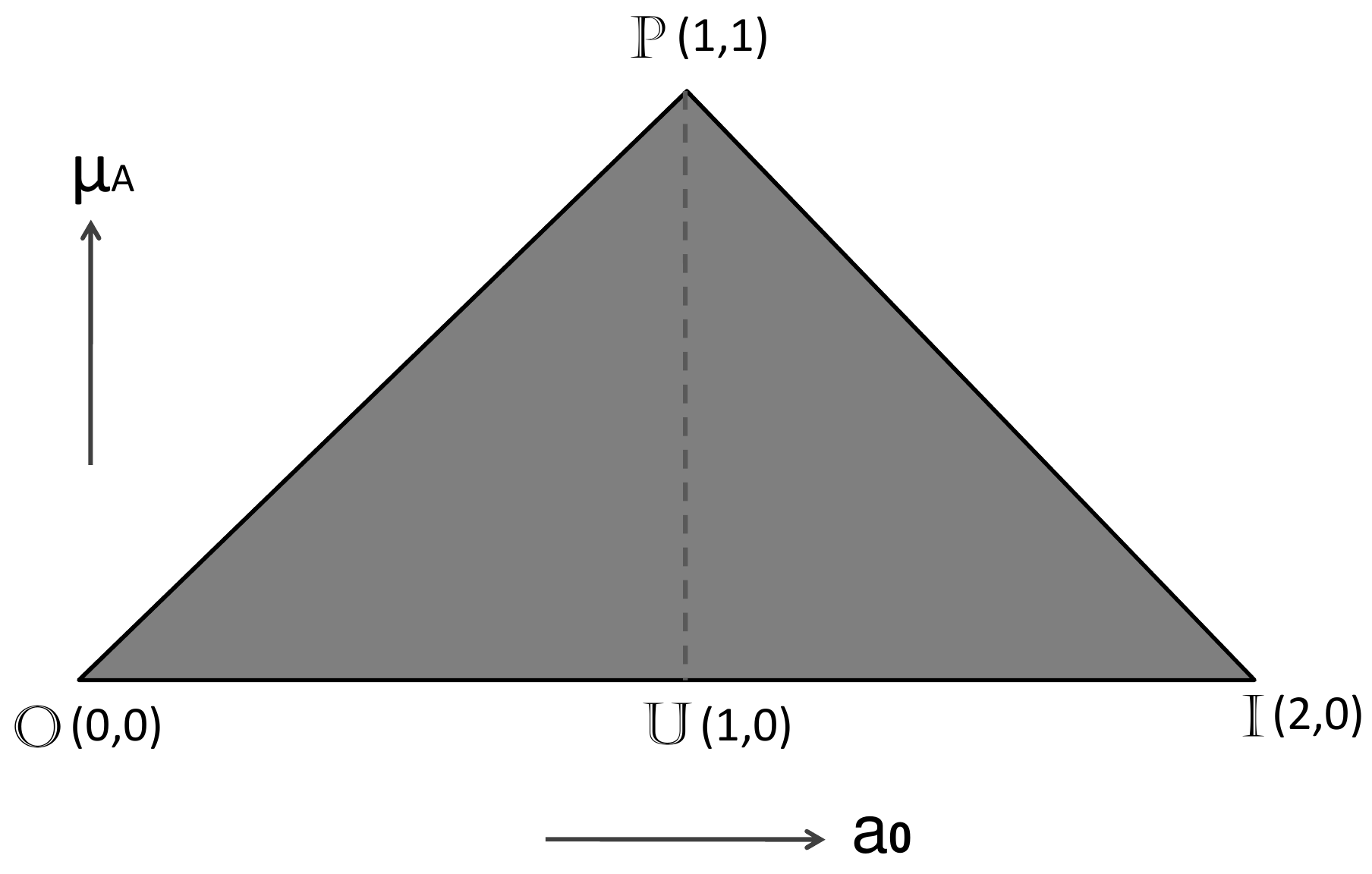}}
\resizebox{4.0cm}{2.5cm}{\includegraphics{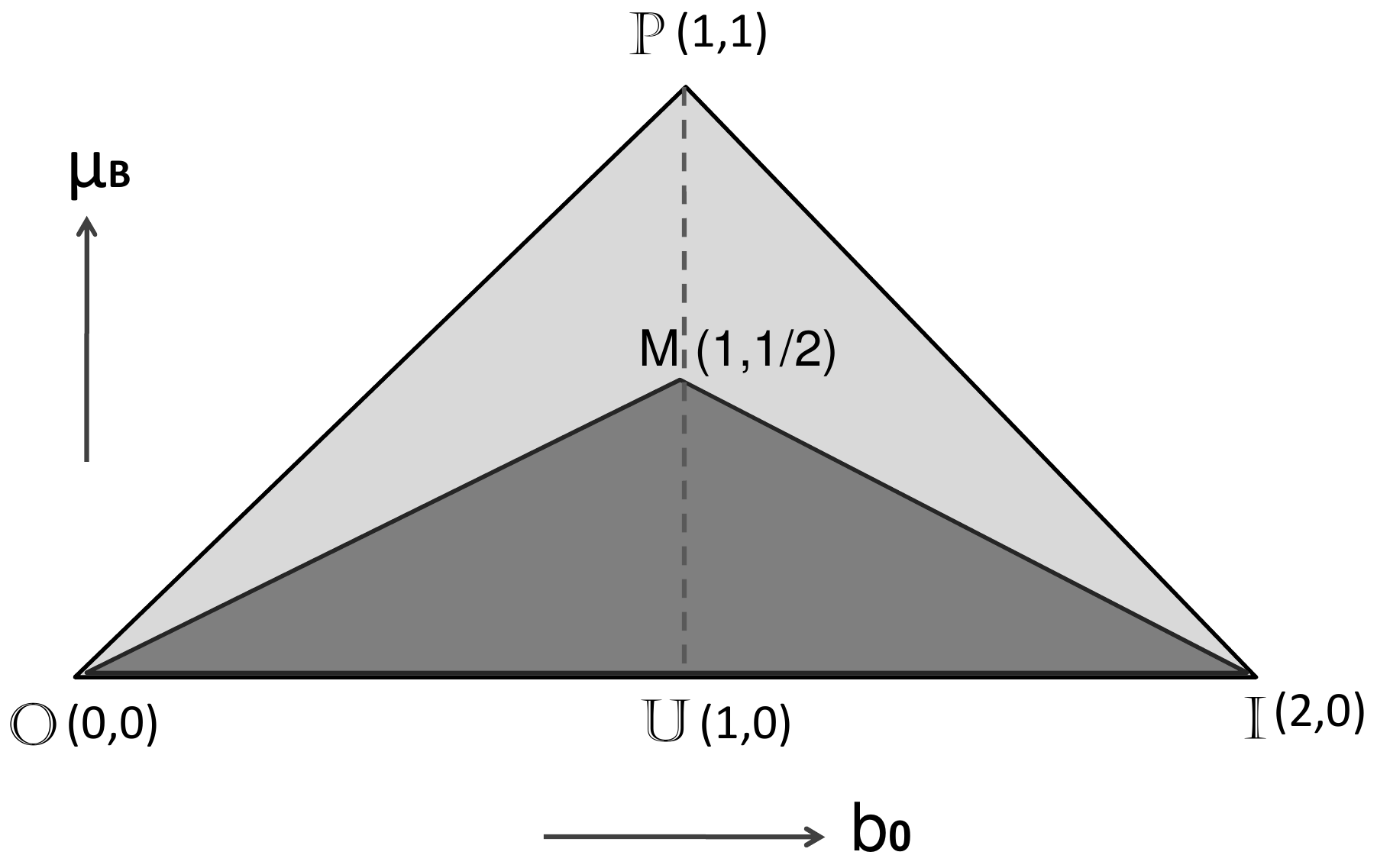}}
\end{array}$
\end{center}
\caption{Alice's (Bob's) parameters can take values from the dark gray triangular region on left (right). Alice's parameters $a_0$ and $\mu_A$ can take any possible value, but Bob's parameters $b_0$ and $\mu_B$ are restricted to come from the region $\mathbb{MOI}$. \label{fig2}}
\end{figure}

Let, Bob's observable $E_B(b_0,\mu_B,\hat{b})$ satisfy restriction $\mu_B\leq \frac{1}{2}\mbox{min}\{b_0,2-b_0\}$ but there is no restriction on Alice's observables, see Fig(\ref{fig2}).

Alice, for her observable $E_A[a_0,\mu_A,\hat{a}]$, declares `$yes$' with
a probability
\begin{equation}
P^{A}_{\hat{\lambda}}(yes)=\frac{a_0}{2} +\frac{1}{2}\mu_A \cos\alpha \label{eq5}
\end{equation}
where $\alpha $ is angle between direction $\hat{a}$ and $\hat{\lambda}$.
On the other hand, for observable $E_B[b_0,\mu_B,\hat{b}]$, Bob declares
`$yes$' with a probability
\begin{equation}
P^{B}_{\hat{\lambda}}(yes)= \frac{b_0}{2}-\mu_B~\mbox{sgn}(\cos\beta) \label{eq6}
\end{equation}
where $\beta$ is the angle between the direction $\hat{b}$ and $\hat{\lambda}$, and $\mbox{sgn}(x)=+1~(-1)~\mbox{for}~x\geq0~(x<0)$.

The joint probability of the outcome ($yes,~yes$), can be calculated from
\begin{equation}
P^{AB}_{lhv}(yes,yes)=\int \rho(\hat{\lambda})~P^{A}_{\hat{\lambda}}(yes)~P^{B}_{\hat{\lambda}}(yes)~d\hat{\lambda} \label{eq7}
\end{equation}
where $\rho(\hat{\lambda})$ is the considered (uniform) distribution of the hidden variable $\hat{\lambda}$.
Evaluating the above integral gives
\begin{equation}
P^{AB}_{lhv}(yes,yes)=\frac{1}{4}[a_0 b_0-\mu_A\mu_B \hat{a}\cdot\hat{b}] \label{eq8}
\end{equation}
which exactly matches with the quantum mechanical prediction for the outcome ($yes,~yes$). The desired quantum mechanical probabilities for the other possible outcomes easily follows, for example, $P^{AB}_{lhv}(yes,no)$ is obtained simply by  replacement $P^{B}_{\hat{\lambda}}(yes)\rightarrow P^{B}_{\hat{\lambda}}(no) = 1-P^{B}_{\hat{\lambda}}(yes) $ in the integrand of the Eq.(\ref{eq7}).

\subsubsection{A fully symmetric model $\mathbb{M}_{fs}$:}
\begin{figure}[b]
\begin{center}$
\begin{array}{cc}
\resizebox{4.0cm}{2.5cm}{\includegraphics{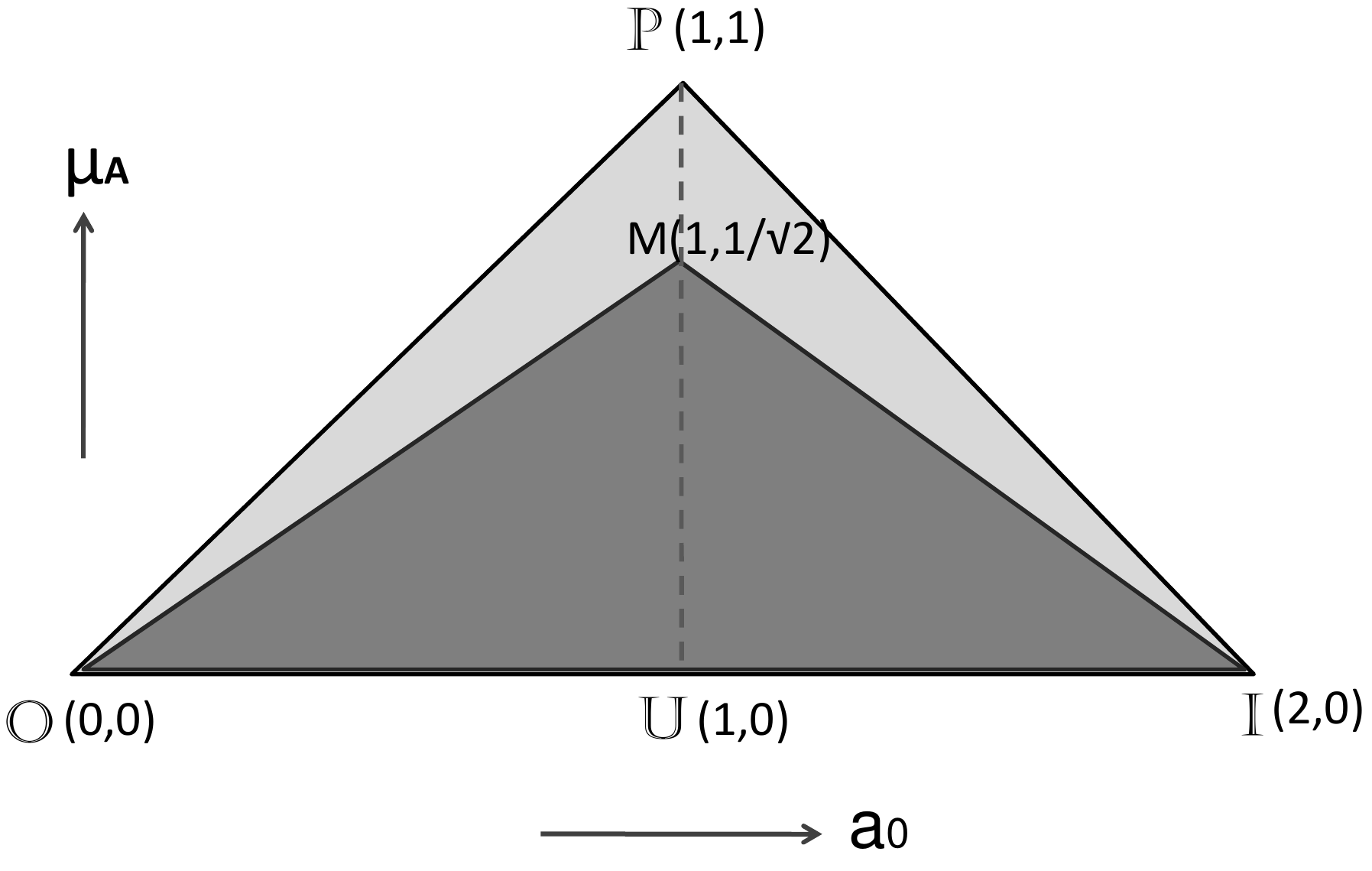}}
\resizebox{4.0cm}{2.5cm}{\includegraphics{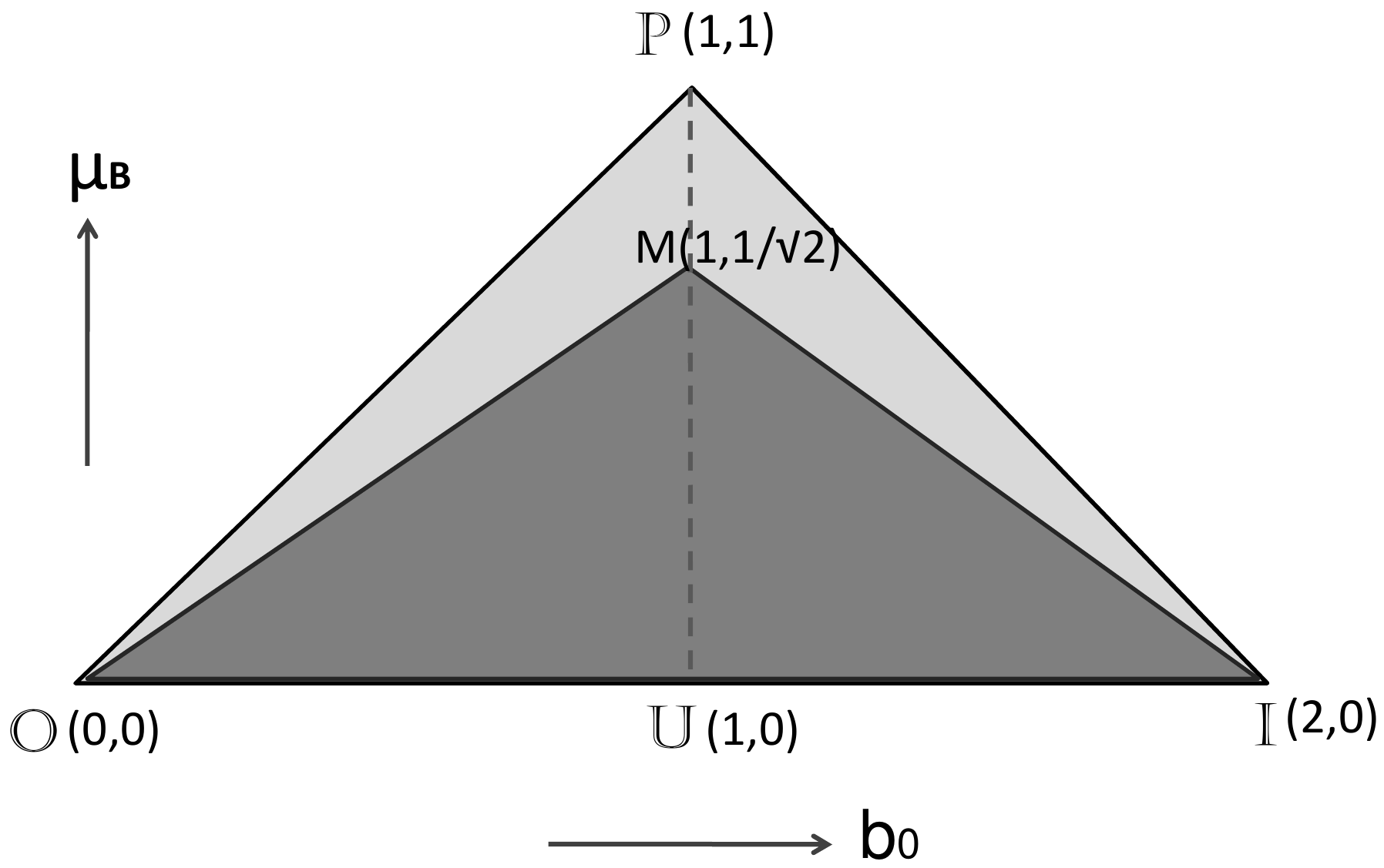}}
\end{array}$
\end{center}
\caption{ Alice's (Bob's) parameters can take values from the dark gray triangular region on left (right). Alice's parameters $a_0$ and $\mu_A$ as well as Bob's parameters $b_0$ and $\mu_B$ are restricted in the same way and come from the triangular region $\mathbb{MOI}$ on the left and right respectively. \label{fig3}}
\end{figure}

Let Alice's and Bob's observables satisfy following restriction (see Fig.(\ref{fig3}))
\begin{eqnarray}
\mu_A\leq \frac{1}{\sqrt{2}}\mbox{min}\{a_0,2-a_0\} \nonumber \\
\mu_B\leq \frac{1}{\sqrt{2}}\mbox{min}\{b_0,2-b_0\} \nonumber
\end{eqnarray}

Alice declares `$yes$', for her observable $E_A[a_0,\mu_A,\hat{a}]$, with a probability
\begin{equation}
P^{A}_{\hat{\lambda}}(yes)=\frac{a_0}{2} +\frac{1}{\sqrt{2}}\mu_A \cos\alpha \label{eq9}
\end{equation}
where $\alpha $ is angle between direction $\hat{a}$ and $\hat{\lambda}$.

Where else Bob declares `$yes$', for his observable $E_B[b_0,\mu_B,\hat{b}]$, with a probability
\begin{equation}
P^{B}_{\hat{\lambda}}(yes)= \frac{b_0}{2} -\frac{1}{\sqrt{2}}\mu_B~\mbox{sgn}(\cos\beta) \label{eq10}
\end{equation}
where $\beta$ is the angle between the direction $\hat{b}$ and $\hat{\lambda}$, and $\mbox{sgn}(x)=+1~(-1)~\mbox{for}~x\geq0~(x<0)$. Like in the fully biased model $ \mathbb{M}_{fb} $, we find that this model ($\mathbb{M}_{fs}$) also simulates the correct statistics for the singlet.

 If we consider unsharp spin properties on both sides with uniform value of  $\mu_A$ and $\mu_B$ and also assume any pair are jointly measurable \cite{busch,gkar} on both sides, then the conditions of the model $\mathbb{M}_{fs}$ are automatically satisfied and hence this LHV model $\mathbb{M}_{fs}$ can simulate the singlet statistics for any arbitrary pair of respective directions $\hat{a}$ and $\hat{b}$ for Alice and Bob.

\subsection{Measure of restriction on observable}
By considering that observables of Alice and Bob are picked from a uniform distribution of all possible two-outcome POVMs, we can define a measure $r$ for restriction on the observables of any of the two parties in the following way. (see Fig.(\ref{fig2}) and Fig.(\ref{fig3}))
\begin{eqnarray}
r
=  \left[1-\frac{\mbox{Area}~(\mathbb{MOI})}{\mbox{Area}~(\mathbb{POI})}\right]\times 100 \label{eq11}
\end{eqnarray}
Now, one can easily calculate that in the model $\mathbb{M}_{fb}$ ($\mathbb{M}_{fs}$), there is $0 \%$ ($29.3 \%$) restriction on Alice's observables wherelse Bob's observables are restricted by $50 \%$ ($29.3 \%$).

Another interesting observation is that model $\mathbb{M}_{fb}$ and $\mathbb{M}_{fs}$ belong to a general class of LHV models, $\{\mathbb{M_{\kappa}}: \kappa \geq 0\}$.

Under the restrictions,
\begin{eqnarray}
\mu_A\leq \kappa~\mbox{min}\{a_0,2-a_0\} \label{eq12}\\
\mu_B\leq \frac{1}{2\kappa}~\mbox{min}\{b_0,2-b_0\} \label{eq13}
\end{eqnarray}
Alice declares ``\emph{yes}'' with a probability
\begin{equation}
P^{A}_{\hat{\lambda}}(yes)=\frac{a_0}{2} +\frac{1}{2\kappa} \mu_A \cos\alpha \label{eq14}
\end{equation}
Bob declares `$yes$' with a probability
\begin{equation}
P^{B}_{\hat{\lambda}}(yes)= \frac{b_0}{2} -\kappa \mu_B~\mbox{sgn}(\cos\beta)\label{eq15}
\end{equation}

\begin{figure}[!ht]
\begin{center}
\resizebox{8.0cm}{6.0cm}{\includegraphics{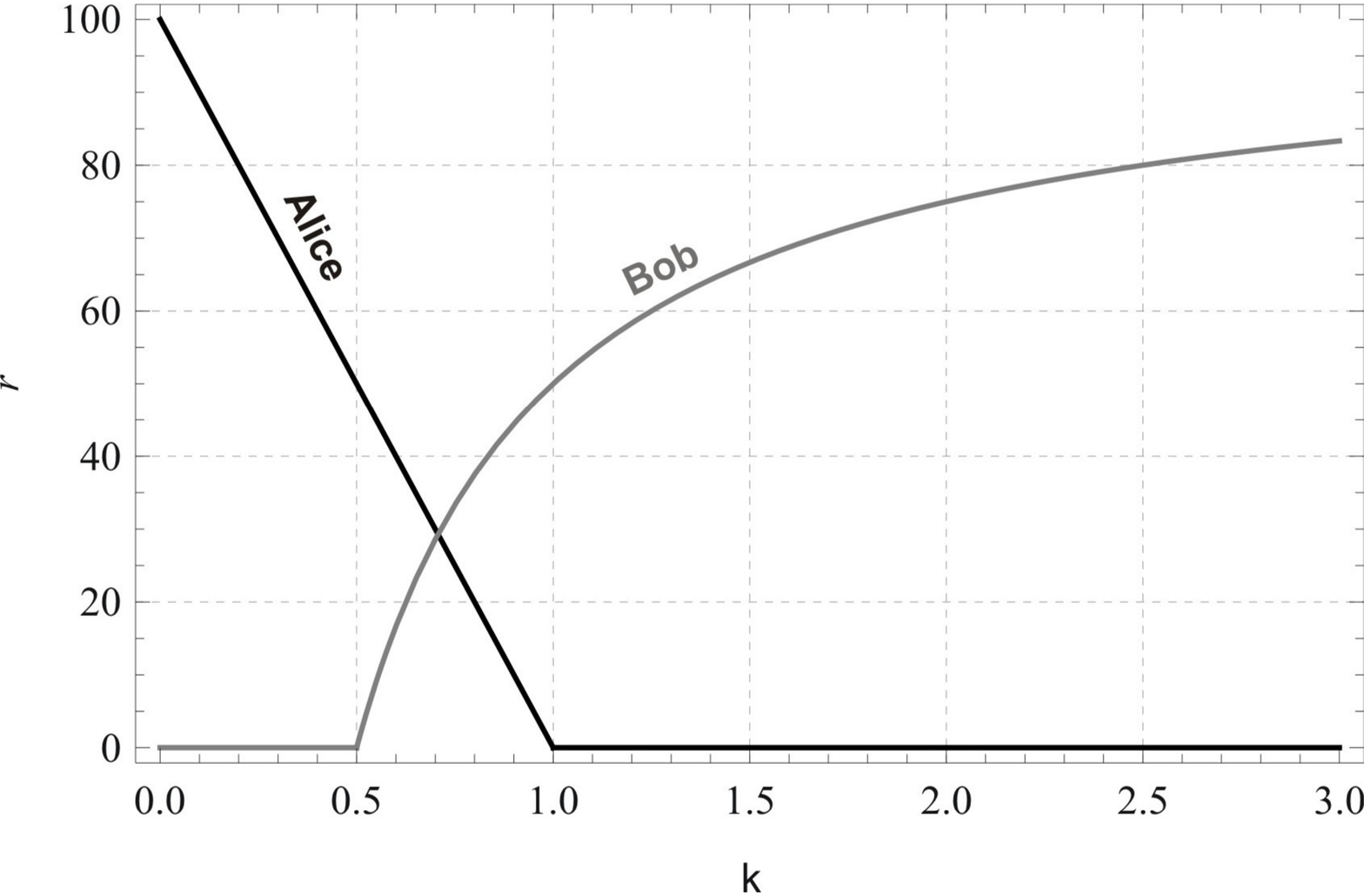}}
\end{center}
\caption{Black (Gray) curve show percentage restriction on Alice's (Bob's) observable for LHV models corresponding to different values of $\kappa$. The intersection point of two curves at $\kappa=\frac{1}{\sqrt{2}}$ correspond to the completly symmetric LHV model $\mathbb{M}_{fs}$. Fully biased model $\mathbb{M}_{fb}$ ($\mathbb{M'}_{fb}$) correspond to $\kappa=\frac{1}{2}$ ($\kappa=1$). All the models for $\kappa \in (0,\frac{1}{2}]\cup [1,\infty)$ are fully biased. \label{fig4}}
\end{figure}

For any nonnegative value of $\kappa$ we get a LHV model---$\mathbb{M}_{fb}$ ($\mathbb{M}_{fs}$) correspond to $\kappa=1$ ($\kappa=\frac{1}{\sqrt{2}}$). In Fig(\ref{fig4}) the two curves show the $\%$ restriction on Alice's and Bob's observable for LHV models corresponding to different values of $\kappa$. The intersection point of two curves correspond to the symmetric model $\mathbb{M}_{fs}$. Observe that $\kappa=\frac{1}{2}$ correspond to another fully biased model, say $\mathbb{M'}_{fb}$, which is same as $\mathbb{M}_{fb}$ except that conditions on Alice's and Bob's observables are interchanged. In fact all the models for which $\kappa \in (0,\frac{1}{2}]\cup [1,\infty)$ are fully biased models, however one can immediately observe that within this subclass, either $\mathbb{M}_{fb}$ or $\mathbb{M'}_{fb}$ is sufficient to simulate any other fully biased model. Thus only the subclass $\{\mathbb{M}_{\kappa}: \kappa \in [1/2, 1]\}$ contains tight LHV models in a sense that they can capture any varying degree of restrictions on Alice's and Bob's observables.

\subsection{A different class of model}
In previous cases, we put restriction on the observable separately on both sides. Now we put the following restriction
on the observable where one of the restrictions involves parameters of both sides;
 \begin{eqnarray}
\frac{1}{\eta}\leq a_0 \leq 2-\frac{1}{\eta}  \label{cond1}\\
\mu_A\mu_B\leq \frac{1}{2\eta}~ \mbox{min} \{b_0, 2-b_0\} \label{cond2}
\end{eqnarray}
where $\eta\geq1$.
Alice and Bob now can simulate singlet statistics according to the following protocol
\begin{equation}
P^{A}_{\hat{\lambda}}(yes)=\frac{a_0}{2} +\frac{1}{2\eta} \cos\alpha
\end{equation}
\begin{equation}
P^{B}_{\hat{\lambda}}(yes)= \frac{b_0}{2} -\eta \mu_A \mu_B~\mbox{sgn}(\cos\beta)
\end{equation}

But this model is obviously non-local as Bob's output involves parameters of observable on both sides. But the model can be made local for a given $\eta$ and fixed $\mu_A$. In this case there is no restriction on direction $\hat{a}$ of Alice's POVM. It might seem that by increasing the value of $\eta$ the range of $a_0$ can be extended but then due to the condition (\ref{cond2}), the range of $\mu_B$ is also restricted accordingly. So in some sense in this model $a_0$ and $\mu_B$ maintain a complementary relation for a given $b_0$.\\

\section{Conclusion}
Simulation of quantum statistics for Werner state by LHV has been an interesting area for understanding the physics of
entanglement \cite{gpp,ps,we,bar}. We have studied  the cases where LHV simulation is possible for singlet state. We find the optimal set of  two outcomes observable for which singlet simulation by LHV is possible under the suggested protocol. It is also interesting that for uniform unsharp parameter, the joint measurability of unsharp spin property on both sides implies LHV model for singlet. It will be interesting to study whether the set can be enlarged with respect to different LHV model.

\begin{acknowledgments}
It is a pleasure to thank Guruprasad Kar and Ramij Rahaman for helpful discussions. SD and AR acknowledge support
from DST project SR/S2/PU-16/2007.
\end{acknowledgments}



\begin{thebibliography}{99}
\bibitem{bell} J.S. Bell, Physics {\bf 1}, 195 (1964); J. F. Clauser, M.A. Horne, A. Shimony and R. A. Holt, Phys. Rev. Lett. {\bf 23}, 880 (1969).
\bibitem{tsb}B. S. Cirel'son, Lett. Math. Phys. 4, 93 (1980).
\bibitem{gpp} N. Gisin, Phys. Lett. A 154, 201 (1991); N. Gisin and A. Peres, Phys. Lett. A 162, 15-17 (1992); S. Popescu and D. Rohrlich, Phys. Lett. A 166, 293 (1992).
\bibitem{ps} Sandu Popescu, Phys. Rev. Lett {\bf 72}, 797 (1994).
\bibitem{we} R. F. Werner, Phys. Rev. A {\bf 40}, 4277 (1989).
\bibitem{acin} G. Toth and A. Acin, Phys. Rev. A {\bf 74}, 030306(R) (2006).
\bibitem{geno} M. Genovese, Physics Reports, {\bf 413}, 319 (2005).
\bibitem{tb} B. F. Toner and D. Bacon, Phys. Rev. Lett. {\bf 91}, 187904 (2003).
\bibitem{cerf} N. J. Cerf, N. Gisin, S. Massar, and S. Popescu, Phys. Rev. Lett. {\bf 94}, 220403 (2005).
\bibitem{gkar1} G. Kar \emph{et al.}, J. Phys. A: Math. Theor. {\bf 44}, 152002 (2011).
\bibitem{hall1} M.J.W. Hall,  Phys. Rev. A  {\bf 82}, 062117 (2010).
\bibitem{deggo} J. Degorre, S. Laplante, and J. Roland, Phys. Rev. A {\bf 72}, 062314 (2005); J. Degorre, S. Laplante, and J. Roland, Phys. Rev. A {\bf 75}, 012309 (2007).
\bibitem{review} H. Buhrman, R. Cleve, S. Massar, and R. de Wolf, Rev. Mod. Phys {\bf 82}, 665 (2010).
\bibitem{hall2} M. J. W. Hall, Phys. Rev. Lett. {\bf 105}, 250404 (2010).
\bibitem{gisin} J. Barrett and N. Gisin, Phys. Rev. Lett. {\bf 106}, 100406 (2011).
\bibitem{hall3} M.J.W. Hall,  Phys. Rev. A  {\bf 84}, 022102 (2011).
\bibitem{manik} M. Banik, MD. R. Gazi, S. Das, A. Rai, and S. Kunkri, J. Phys. A: Math. Theor. {\bf 45}, 205301 (2012).
\bibitem{bru} N. Brunner, N. Gisin, S. Popescu and V. Scarani, Phys. Rev A  {\bf 78}, 052111 (2008).
\bibitem{ban} J. D. Banacal, C. Branciard, and N. Gisin, Adv. Math. Phys. {\bf 2010}, 293245 (2010).
\bibitem{bar} J. Barrett, Phys. Rev. A 65, 042302 (2002).
\bibitem{bu} P. Busch, M. Grabowski and P. Lahti, Operational Quantum
Physics, Springer, Berlin, 1995/1997; P. Busch, P. Lahti, P. Mittelstaedt, The Quantum Theory
of Measurement, Springer, Berlin, 1991, 2nd. ed. 1996.
\bibitem{busch} P. Busch, Phys. Rev. D {\bf 33 }, 2253 (1986).
\bibitem{gkar} G. Kar and S Roy, Rivista del Nuovo Cimento {\bf 22}, 1 (1999).
\bibitem{busch1} P. Busch, Found. Phys.{\bf 17 }, 905 (1987).
\bibitem{kraus} K. Kraus, States, Effects and Operations, Lecture Notes in Physics, vol. 190, Springer, Berlin, 1983.

\end{thebibliography}
\end{document}